\documentclass[pre,twocolumn,superscriptaddress,citeautoscript]{revtex4-1}

\usepackage{epsfig}
\usepackage[usenames, dvipsnames]{color}
\usepackage{natbib}
\usepackage{xifthen}
\usepackage{bm}
\usepackage[version=3]{mhchem} % Formula subscripts using \ce{}
\usepackage{mciteplus}
\usepackage{multirow}
\usepackage{graphicx}% Include figure files
\usepackage{dcolumn}% Align table columns on decimal point
\usepackage{bm}% bold math
\usepackage{mathptmx,amsmath,amssymb,txfonts}
\usepackage{subfig}
\usepackage{blindtext}
\newcommand{\vect}[1]{\boldsymbol{#1}}
\begin{document}
\title{Instability of the Body-Centered Cubic Lattice within the Sticky Hard Sphere and Lennard-Jones Model obtained from Exact Lattice Summations}
\date{\today}

\author{Antony Burrows}
\affiliation{Centre for Theoretical Chemistry and Physics, The New Zealand Institute for Advanced Study, Massey University Auckland, Private Bag 102904, 0632 Auckland, New Zealand}
\author{Shaun Cooper}
\affiliation{School of Natural and Computational Sciences, Massey University Auckland, Private Bag 102904, 0632 Auckland, New Zealand.}
\author{Peter Schwerdtfeger}
\email[Corresponding author, email:]{peter.schwerdtfeger@gmail.com}
\affiliation{Centre for Theoretical Chemistry and Physics, The New Zealand Institute for Advanced Study, Massey University Auckland, Private Bag 102904, 0632 Auckland, New Zealand}

\begin{abstract}
A smooth path of rearrangement from the body-centered cubic (bcc) to the face-centered cubic (fcc) lattice is obtained by introducing a single parameter to cuboidal lattice vectors. As a result, we obtain analytical expressions in terms of lattice sums for the cohesive energy. This is described by a Lennard-Jones (LJ) interaction potential and the sticky hard sphere (SHS) model with an $r^{-n}$ long-range attractive term. These lattice sums are evaluated to computer precision by expansions in terms of a fast converging series of Bessel functions. Applying the whole range of lattice parameters for the SHS and LJ potentials demonstrates that the bcc phase is unstable (or at best metastable) toward distortion into the fcc phase. Even if more accurate potentials are used, such as the extended LJ potential for argon or chromium, the bcc phase remains unstable. This strongly indicates that the appearance of a low temperature bcc phase for several elements in the periodic table is due to higher than two-body forces in atomic interactions.
\end{abstract}
\maketitle

\section{Introduction}
The stability of different bulk phases and their possible connections through distortions and rearrangements in phase transitions remain an open and challenging field in solid-state physics \cite{young1991}. Solid-to-solid phase transitions are commonly modeled by computer intensive molecular dynamic or Monte-Carlo simulations at finite temperatures and pressures \cite{Binder1985,Gomez2019}, or by various algorithms to find phase transition paths on a Born-Oppenheimer hypersurface \cite{Caspersen2005}. For example, the relative stability of the fcc versus the hexagonal close packing (hcp) and possible transition mechanisms between these two phases for the rare gas elements has been a matter of a long-standing controversy \cite{Mau1999,Stillinger2001,Krainyukova-2012,Schwerdtfeger-2016,Bingxi2017,Wiebe2020,Smits2020a}. While fcc has a higher excess entropy compared to hcp by a rather small difference (for the hard sphere model it is 0.00115$\pm$0.00004 $k_B$ per sphere \cite{Mau1999}), the energetic stability of the fcc over the hcp phase for the rare gas solid argon (at low temperatures and pressures) is due to quantum effects (phonon dispersion) \cite{Schwerdtfeger-2016,Wiebe2020}. Similarly, the transformation between the bcc$\leftrightarrow$fcc phases has been the subject of many discussions as the exact martensitic type of transformation path for a solid, such as in iron-based materials, or in clusters, is still being debated \cite{Kraft1993,Rollmann2007,Cayron2015}.

It is commonly believed that strong repulsive forces favor close-packed arrangements such as fcc or hcp, whereas soft repulsion favors less dense packed structures such as bcc \cite{Agrawal1995,Prestipino2005,Likos2007,Bharadwaj2017}. Laird showed that the bcc phase is unstable within the hard-sphere model \cite{Laird1992}, while Hoover et al. and later Agrawi and Kofke showed that soft repulsive potentials of the form $ar^{-n}$ with small $n$ values are required to stabilize the bcc phase \cite{HooverGrover1972,Agrawal1995}. Very recently Ono and Ito used phonon dispersion curves to show that soft Lennard-Jones forces are required to turn the bcc phase into a minimum \cite{Ono2021}. However, as minima can be very shallow on a energy hypersurface, one requires accurate numerical or analytical methods to determine if the bcc phase represents a (metastable) minimum for a two-body potential or not. Moreover, inverse power law potentials such as the LJ potential have the advantage that properties such as the cohesive energy can be evaluated analytically through lattice sums \cite{borwein-2013,burrows-2020}. If a single path through a lattice parameter can be found \cite{Caspersen2005} describing smoothly the bcc$\leftrightarrow$fcc transition (not necessarily a minimum energy path), one gains valuable insight into the stability of the bcc phase.

Conway and Sloane introduced the isodual mean-centered cuboidal lattice (mcc) which can be seen as an average between the bcc and the fcc lattice \cite{Conway1994}. They introduced lattice vectors depending on two parameters connecting the bcc, mcc and fcc lattices. Recently we were able to find fast converging lattice sums for these cuboidal lattices derived from their corresponding Gram matrices and quadratic forms \cite{burrows2021a}. These lattice sums, which can be evaluated to computer precision, will be introduced in the next section and applied to analyse the energy profile of the bcc lattice distortion into the fcc densest packing using LJ and SHS interaction potentials. For more realistic two-body forces we apply  extended Lennard-Jones potentials \cite{Schwerdtfeger-2006} for Ar$_2$ \cite{Smits2020a} and Cr$_2$.

\section{Method}
\begin{figure}[htb!]
  \begin{center}
\includegraphics[scale=.23]{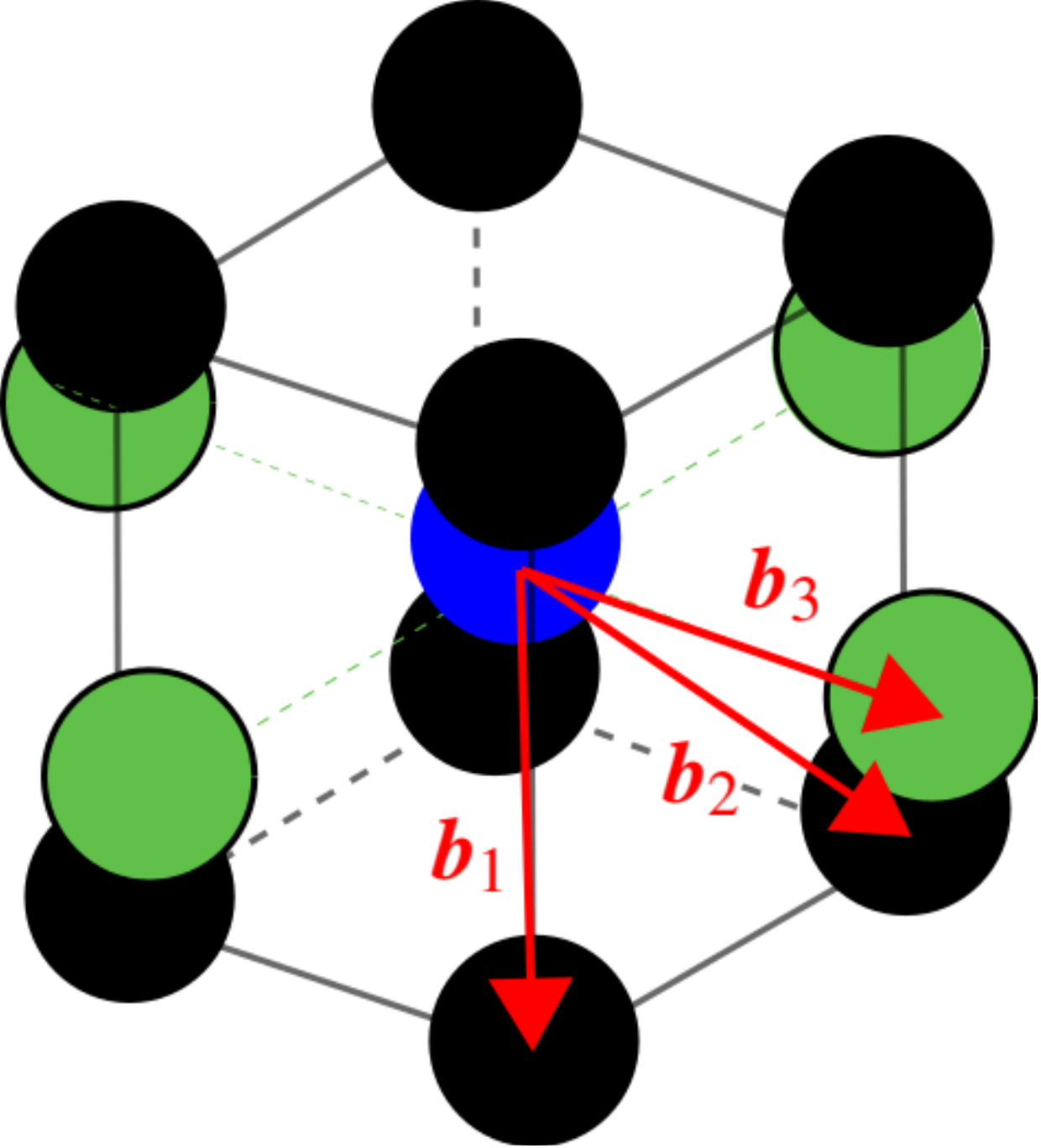}
  \caption{The mcc  ($A=1/\sqrt{2}$) lattice and corresponding non-orthogonal basis vectors (red) according to (\ref{eq:latticevectors}). In green the atoms moving towards the central atom (blue) becoming nearest neighbors in the fcc lattice are shown.}
  \label{fig:lattices}
  \end{center}
  \end{figure}

Lattice vectors for the primitive cell depending on a single parameter $A$ are defined by 
\begin{align}
\label{eq:latticevectors}
\vect{b}_1^\top(A)=\left( 1,0,0 \right) \quad , \quad \vect{b}_2^\top(A)=\left( \frac{A}{A+1},\frac{\sqrt{2A+1}}{A+1},0 \right), \\
\nonumber
\vect{b}_3^\top(A)= \left( \frac{1}{A+1},\frac{1}{(A+1)\sqrt{2A+1}},\sqrt{\frac{4A}{(A+1)(2A+1)}}\right).
\end{align}
The corresponding Gram matrix for the quadratic form is given by the scalar product between these lattice vectors,
\begin{equation}
G_{ij}(A) = \langle\vect{b}_i (A),\vect{b}_j(A)\rangle = \frac{1}{A+1}\begin{pmatrix}
A+1 & A & 1 \\
A & A+1 & 1 \\
1 & 1 & 2
\end{pmatrix}.
\end{equation}
The cuboidal lattices are defined in the range $A\in[\frac{1}{3},1]$ \cite{burrows2021a}, and for the special values of $A=1/3$, $A=1/2$,  $A=1/\sqrt{2}$, and $A=1$ lattice vectors for the acc (axial centered cuboidal \cite{Conway1994}), bcc, mcc and fcc lattices are obtained, with number of nearest neighbors of 10, 8, 8 and 12 respectively. This sets the minimal distance between two lattice points to 1 for the range $A\in[\frac{1}{3},1]$, which ensures that the lattice deformation is compatible with the hard sphere model. The volume spanned by these three vectors is $V(A)=\sqrt{{\rm det}G(A)}=2A^{1/2}(A+1)^{-3/2}$ with a maximum volume at the bcc structure ($A=1/2$). The mcc lattice with the corresponding lattice vectors (\ref{eq:latticevectors}) is shown in Figure \ref{fig:lattices}.
  
The choice of the basis vectors (\ref{eq:latticevectors}) has the advantage that only $\vect{b}_2$ and $\vect{b}_3$ move in this 3D lattice transformation. The length of $\vect{b}_1$ and $\vect{b}_2$ is 1 for all $A$ values considered, and the angle between $\vect{b}_1$ and $\vect{b}_3$ is the same as between  $\vect{b}_2$ and $\vect{b}_3$. From the Gram matrix one obtains the atomic packing fraction or packing density \cite{conway2013sphere} for the cuboidal lattices as $\rho(A) =\frac{\pi}{12}\sqrt{\frac{(A+1)^3}{A}}$ \cite{burrows2021a}. This yields the well known values for fcc ($\rho(1) =\pi\sqrt{2}/6$) and bcc ($\rho(\frac{1}{2}) =\pi\sqrt{3}/8$). In fact, from this formula we deduce that bcc is the least packed arrangement of all the cuboidal lattices considered here. 

Using an $(a,b)$ LJ potential in its most general form \cite{Gruneisen1912,Jones-1925}
\begin{equation} \label{eq:LJ}
V_{\rm LJ}(r,a,b)=\frac{ab}{a-b}\epsilon \left[ \frac{1}{a}\left( \frac{r_e}{r} \right)^{a} - \frac{1}{b} \left( \frac{r_e}{r} \right)^{b} \right] ,
\end{equation}
where $r_e$ is the minimum (equilibrium) distance, $\epsilon>0$ is the dissociation energy and $a>b>3$ are real numbers. We obtain an analytical expression for the cohesive energy in terms of lattice sums $L(a,A)$ and the nearest neighbor distance $R$ in the lattice \cite{burrows-2020},
\begin{equation} \label{eq:CohEnergy}
E_{\rm LJ}(R,a,b,A) =\frac{ab\epsilon}{2(a-b)}\left[\frac{1}{a} L(a,A)\left(\frac{r_e}{R}\right)^{a} - \frac{1}{b} L(b,A)\left(\frac{r_e}{R}\right)^{b}\right] .
\end{equation}
Here, $b>3$ is required to avoid the singularity in $L(b,A)$ at $b=3$ \cite{Schwerdtfeger-2006} (although these lattice sums can be analytically continued \cite{borwein-2013,borwein1998convergence,burrows2021a}). The lattice sums $L(a,A)$ are defined through their corresponding quadratic forms $\vect{i}^\top G \vect{i}~,~ \vect{i}\in\mathbb{Z}^3$ by \cite{conway2013sphere}

\begin{equation}\label{eq:LttSum}
L(a,A) = {\sum_{\vect{i}\in\mathbb{Z}^3}}^{\prime} \left(\frac{1}{\vect{i}^\top G \vect{i}}\right)^{a/2}= {\sum_{i,j,k}}^{\prime} \left(\frac{A+1}{A(i+j)^2+(j+k)^2+(i+k)^2}\right)^{a/2} ,
\end{equation}
where the prime symbol indicates that the term corresponding to $\vect{i}^\top=(0,0,0)$ is omitted in the summation. For small values of $a$, these triple sums are slowly convergent and one needs to find expansions in terms of fast converging series to obtain computer precision \cite{borwein-2013}. A number of methods to achieve this have recently been introduced by our group \cite{burrows-2020,burrows2021a}. A program to evaluate these lattice sums including the cuboidal lattices considered here is freely available from our website \cite{ProgramJones}. For this work we use either the Terras decomposition of the Epstein zeta function \cite{terras-1973,burrows-2020} or the decomposition in terms of Jacobi $\theta$ functions and integral transforms to produce series expansions in terms of Bessel functions \cite{burrows-2020,burrows2021a}.

The SHS model can easily be obtained in the limit of $a\rightarrow\infty$ of the LJ potential \cite{Trombach2018}, and the cohesive energy given by the expression
\begin{equation}\label{eq:SHS}
E_{\rm SHS}(R,b,A) = \lim_{a\rightarrow\infty} E_{\rm LJ}(R,a,b,A) = -\frac{\epsilon}{2} L(b,A)\left(\frac{r_e}{R}\right)^{b},
\end{equation}
with $R\ge r_e$. This gives a direct relation between the SHS energy of the solid and the corresponding lattice sum.

It is convenient to introduce dimensionless units ($R^*=R/r_e$ and $E^*=E/\epsilon$). The minimum nearest neighbor distance for a cuboidal lattice can be found from (\ref{eq:CohEnergy}),
\begin{equation} \label{eq:mindist}
 R^*_{\rm min}(a,b,A)=\left[ \frac{L(a,A)}{L(b,A)}\right]^{\tfrac{1}{a-b}} .
\end{equation}
For the SHS model this reduces to $R^*_{\rm min}=1$. The cohesive energy at minimum becomes
\begin{align} \label{eq:minenergy}
E^*&(R^*_{\rm min},a,b,A)=-\frac{1}{2}\left[ \frac{L(b,A)^a}{L(a,A)^b}\right]^{\tfrac{1}{a-b}} ,
\end{align}
and for the SHS model we attain $E^*(R^*_{\rm min}=1;b,A)=-L(b,A)/2$. Finally, a more realistic two-body potential is used, where lattice sum techniques can still be applied. This requirement is fulfilled by the extended Lennard-Jones (ELJ) potential, which is an inverse power series expansion in terms of the distance $R$, 
 \begin{equation} \label{eq:extLJ}
E_{\rm ELJ}(R,c_n,A) =\frac{1}{2}\sum_{n=1}^{n_\textrm{max}} c_nL(a_n,A)R^{-a_n} ,
\end{equation}
 with $\sum_n c_n=-\epsilon$ and $a_n>3$ \cite{Schwerdtfeger-2006,Smits2020a}.

 \section{Results}
\begin{figure}[ht!]
  \begin{center}
 \includegraphics[scale=0.22]{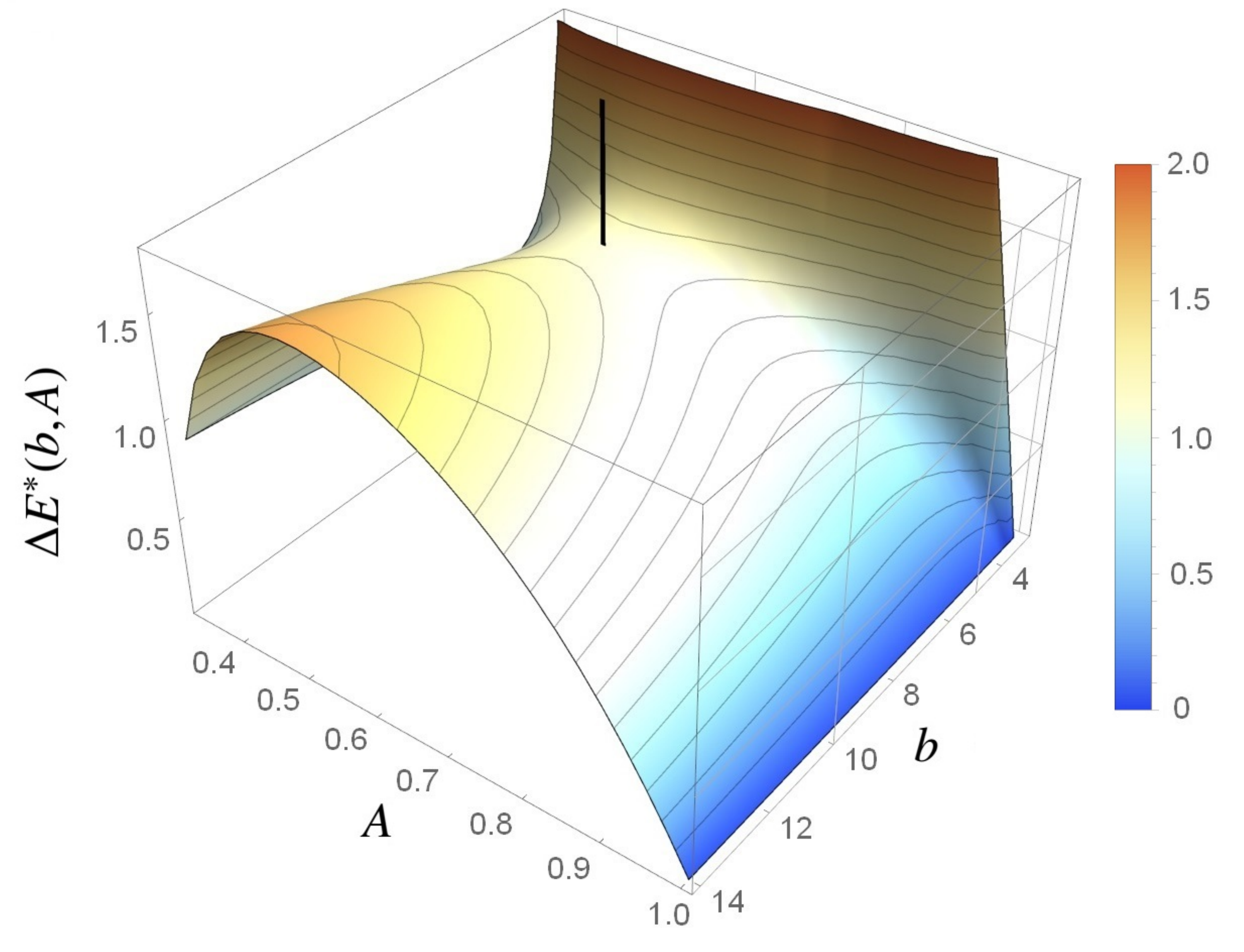}
  \caption{Difference in cohesive energies $\Delta  E^*(b,A) = \tfrac{1}{2}\left[ L(b,A=1)-L(b,A)\right]$ between the cuboidal lattices and fcc for various exponents $b$ and lattice parameter $A$ of the SHS model. Contour interval chosen is 0.125. The vertical black line at $b=5.493634$ shows the point of least instability for the bcc lattice.}
  \label{fig:SHS}
  \end{center}
  \end{figure}

Starting with the discussion of the SHS model, the difference in cohesive energies between the $A$-dependent cuboidal lattices and the fcc lattice ($A=1$) as a function of the two parameters $b$ and $A$,
 $\Delta  E^*(b,A) = \left[ L(b,A=1)-L(b,A)\right]/2$ at $R^*_{\rm min}=1.0$, is shown in Figure \ref{fig:SHS}. It is evident that the SHS model predicts a maximum in energy at the bcc structure. In fact, it was proved recently that $\partial L(b,A)/\partial A=0$ and $\partial^2 L(b,A)/\partial A^2>0$ at $A=\frac{1}{2}$ (bcc) for all $b\in (3,\infty)$ \cite{burrows2021a}. Despite the path being chosen along the $A$ parameter, this most likely does not represent the true minimum energy path, it is clearly downhill energetically towards the fcc structure. As a result, the bcc lattice is unstable with respect to distortion to fcc within the SHS model. There is also the opposite path towards the acc crystal ($A=1/3$), which has to our knowledge not yet been observed in nature. Figure \ref{fig:SHS} shows that for low $b$ values, $\Delta  E^*(b,A)$ starts to increase again (at lower exponents $\Delta  E^*(b,A)\rightarrow \infty$ for $b\rightarrow 3$). The most stable bcc lattice is observed at $\Delta  E^*(b=5.49363406\dots,\frac{1}{2}) = 1.090510595\dots$, with a $b$ value close to the exponent $b=6$ used for dispersive type of forces.
  
As the SHS model clearly has its limitations, we turn to the more accurate $(a,b)$ LJ potential, i.e., we introduce softer repulsive walls to the SHS model. This will also remove the discontinuity in the $\Delta  E^*(a,b,A)$ curve at the fcc point ($A=1$). Due to the attractive long-range lattice forces, the minimum distance between two lattice points is $R^*_{\rm min}(a,b,A)<1$ for finite $(a,b)$ exponents provided that  $a>b>3$, see eq.(\ref{eq:mindist}). $R^*_{\rm min}(a,b,A)$ does not vary much with respect to $A$ for a fixed $(a,b)$ combination (see appendix). The minimum distance for the (12,6) LJ potential for the bcc lattice is $R^*_{\rm min}(12,6,\frac{1}{2})$=0.951864819 compared to $R^*_{\rm min}(12,6,1.0)$=0.9712336910 for the fcc lattice. That is, lattice interactions lead to a contraction in the nearest neighbor distance by about 4.8\% for the bcc lattice. A deviation from this behavior is observed only for large $a$ values, as eventually the minimum distance has to approach the SHS limit of $R^*_{\rm min}=$1.0. Here $R^*_{\rm min}$ at bcc turns into a very shallow maximum (see appendix). From eq.(\ref{eq:mindist}) and  $\partial L(b,A)/\partial A$=0 at $A=\frac{1}{2}$  \cite{burrows2021a} it follows that $\partial R^*_{\rm min}(a,b,A)/\partial A=0$ at $A=\frac{1}{2}$, and the bcc point remains a critical point.

\begin{figure}[hbt]
  \begin{center}
  \includegraphics[scale=0.32]{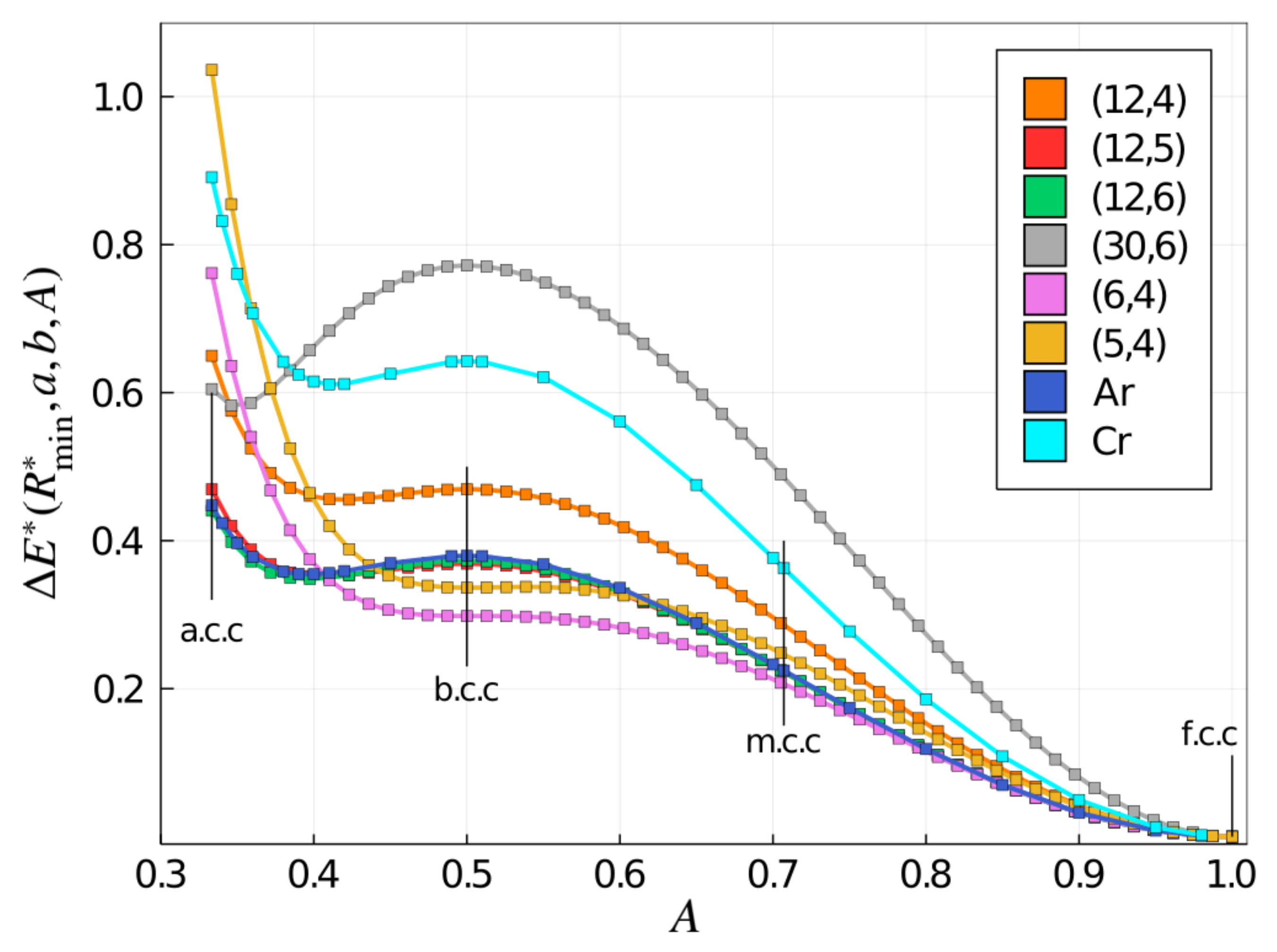}\\
  \caption{Cohesive energy differences $\Delta  E^*(R^*_{\rm min},a,b,A)=E^*(R^*_{\rm min},a,b,A)-E^*(R^*_{\rm min},a,b;A=1)$ for the $(a,b)$ LJ potential dependent on the lattice parameter $A$, and for the two ELJ potentials of argon and chromium (see appendix).}
  \label{fig:Energies}
  \end{center}
  \end{figure}

\begin{figure}[hbt]
  \begin{center}
  \includegraphics[scale=0.27]{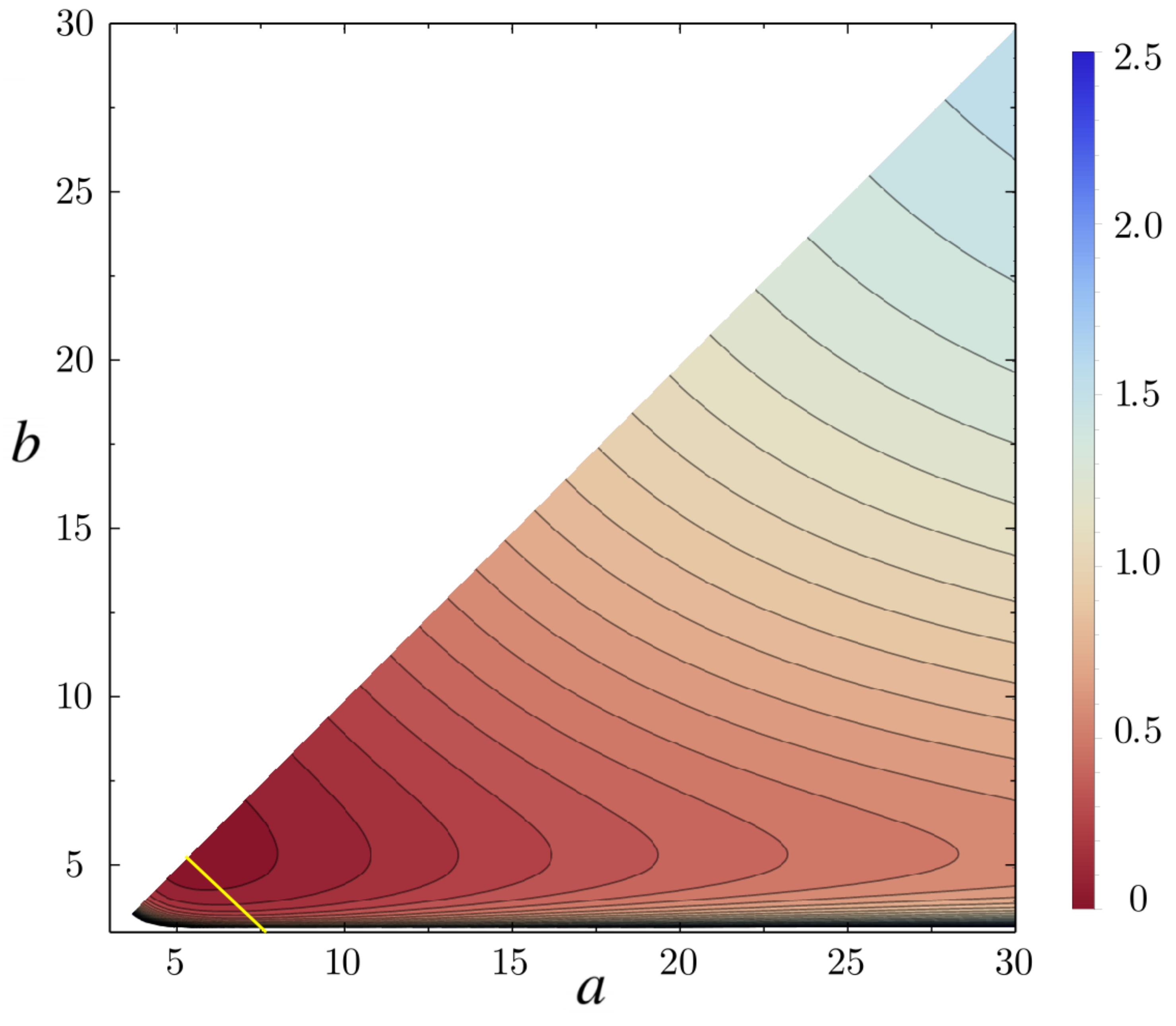}
  \caption{Energy difference $\Delta E^*_{\rm bcc,fcc}(a,b) = E^*(R^*_{\rm min,bcc},a,b,\tfrac{1}{2})-E^*(R^*_{\rm min,fcc},a,b,1)$ between the bcc and fcc lattice for the $(a,b)$ LJ potential. The yellow curve indicates the phase transition line to a metastable bcc state.}
  \label{fig:hypersurface}
  \end{center}
  \end{figure}

Shorter distances are usually associated with greater stability of the lattice which could stabilize the bcc phase. This is however not the case as Figure \ref{fig:Energies} shows. In fact, the bcc lattice is {\it not} a stable lattice for any $(a,b)$ LJ potential, i.e., it will continuously distort by lowering the energy along the distortion parameter $A$  toward the most densely packed and stable fcc lattice.
%\begin{align} \label{eq:minenergydif}
%\nonumber
%\Delta &E^*_{\rm bcc,fcc}(a,b) = E^*(R^*_{\rm min,bcc},a,b,\tfrac{1}{2})-E^*(R^*_{\rm min,fcc},a,b,1)\\
%&=\frac{1}{2}\frac{L(b,1)^\frac{a}{a-b}}{L(a,1)^\frac{b}{a-b}} \left[ 1-\left(\frac{L(a,1)}{L(a,\tfrac{1}{2})}\right)^{\tfrac{b}{a-b}} %\left(\frac{L(b,\tfrac{1}{2})}{L(b,1)}\right)^{\tfrac{a}{a-b}}\right] .
%\end{align}
%and it depends on the expression in the square brackets if bcc is more stable than fcc. 
From eq.(\ref{eq:minenergy}) the energy difference between the bcc and the fcc lattice for a $(a,b)$ LJ potential becomes $\Delta E^*_{\rm bcc,fcc}(a,b) = E^*(R^*_{\rm min,bcc},a,b,\tfrac{1}{2})-E^*(R^*_{\rm min,fcc},a,b,1)$. Figure \ref{fig:hypersurface} shows that $\Delta E^*_{\rm bcc,fcc}(a,b)>0$ for all $a>b>3$, thus bcc remains energetically unstable for a general LJ potential. For a very small $(a,b)$ range, however, the bcc phase becomes metastable, i.e., the minimum at $A<\frac{1}{2}$ shifts toward the bcc structure, see Figure \ref{fig:hypersurface}. The $(a,b)$ phase transition line from the unstable to the metastable bcc lattice is approximately described by the polynomial $a_{\rm PT} = -6.3829845\times 10^{-4}b_{\rm PT}^3 + 3.8186745\times 10^{-2}b_{\rm PT}^2 -1.3466248b_{\rm PT} + 1.1373783\times 10^1$ with $a_{\rm PT}>b_{\rm PT}\in(3,5.25673]$ (see appendix), and we see an almost linear behavior as shown in Figure \ref{fig:hypersurface}. This also explains why Ono and Ito obtained imaginary phonon frequencies for some low $(a,b)$ combinations \cite{Ono2021} (their results have to be taken with some care as the $r^{-3}$ potential used leads to a singularity in the cohesive energy). In fact, the bcc structure becomes metastable if and only if $L(a,A)\partial^2L(a,A)/\partial A^2<L(b,A)\partial^2L(b,A)/\partial A^2$ for $A=\frac{1}{2}$ and $a>b>3$ (see appendix). However, these minima appear at energies $\Delta E^*_{\rm bcc,fcc}(a,b)>$0.2 ($a<7.660388$) for rather unphysical potentials, with low $\Delta E^*$ values only if $a\approx b$. As an example, for a (4,3.1) LJ potential the bcc structure is a minimum at $\Delta E^*_{\rm bcc,fcc}=170.2$ with an activation barrier of ${\Delta E^*}^\#=12.2$ situated at $A=0.6$ on the path toward the distortion to the fcc structure. The distortion along the $A$ parameter can also occur towards a metastable lattice with $A<\frac{1}{2}$ and higher packing density. Alternatively, with a (12,6) LJ potential the metastable minimum sits at a lattice with $A=0.3962483\dots$ and packing density $\rho=0.6861655\dots$. Finally, the mcc lattice is just a lattice along the energetic downward path towards fcc as for the SHS model.
  
As $\partial L(b,A)/\partial A=0$ at $A=\frac{1}{2}$ \cite{burrows2021a} we obtain $\partial E^*(R^*_{\rm min},a,b,A)/\partial A=0$ at $A=\frac{1}{2}$ (see appendix), and the bcc structure remains a critical point for all $(a,b)$ combinations. Moreover, if the exponent $a$ responsible for the repulsive wall increases, we approach the limit of the SHS potential with much higher energies compared to the LJ potential. Here we mention that by applying an inverse power law potential for the repulsive wall (opposed to the long-range part in the SHS model), Agrawal and Kofke showed from Monte Carlo simulations that the bcc phase is unstable \cite{Agrawal1995}.  
  
 The Einstein frequency $\omega_E$ of a single atom moving in the field of all other atoms can be expressed analytically in terms of lattice sums \cite{Smits2020a}. $\omega_E(a,b,A)>0$ for all $A\in[\frac{1}{3},1]$ and $a>b>3$. As a consequence, a single atom is locked and more than one atom has to move simultaneously along the bcc$\rightarrow$fcc path similar to a Zener martensitic transformation \cite{rifkin1984,Cayron2015}. 
  
 The question remains as to why bcc lattices are observed in nature given their instability, large volume and small bulk modulus within the cuboidal structures. It is clear that two-body forces favor dense packings with the largest kissing number for an atom, that is fcc or hcp. The answer therefore lies in the failure of the two-body potential to correctly describe the interactions in the crystal, i.e., neglecting important higher than two-body interactions (and perhaps quantum effects for quantum solids such as helium). It is well known that the many-body expansion is only slowly convergent for metallic systems \cite{Kaplan1996,N-body.AH2007}. 
 %For example, Li$_2$ is bound with a dissociation energy of 1.142 eV \cite{Linton1996}, but in comparison has a rather small cohesive energy of 1.63  eV for the bulk bcc structure \cite{kittel1996} due to higher than two-body forces. 
 To see if the form of the LJ potential limits our conclusion, a more accurate ELJ two-body potential is taken, derived from relativistic coupled cluster theory for argon \cite{CybulskiToczylowski1999,Schwerdtfeger-2016}. As in the case for the (12,6) LJ potential, the ELJ potential has a minimum $R^*_{\rm min}(A)$ value at the bcc structure (see appendix). More importantly, the $E_{\rm ELJ}^*(A)$ curve does not change substantially in shape and is only slightly shifted compared to the (12,6) LJ potential, as shown in Figure \ref{fig:Energies}. This is perhaps expected from the comparison between the two potentials (see appendix), and from the fact that for the fcc structure $E^*(R^*_{\rm min},1.0)=7.8532$  \cite{Smits2020a} for the ELJ potential and close to $E^*(R^*_{\rm min},12,6,1.0)=-L_6^2/(2L_{12})=8.6102$ for the (12,6) LJ potential (exp. $E^*=6.4951$ using the data from Ref.\cite{Schwalbe-1977}).
 \begin{figure}[hb]
  \begin{center}
  \includegraphics[scale=0.3]{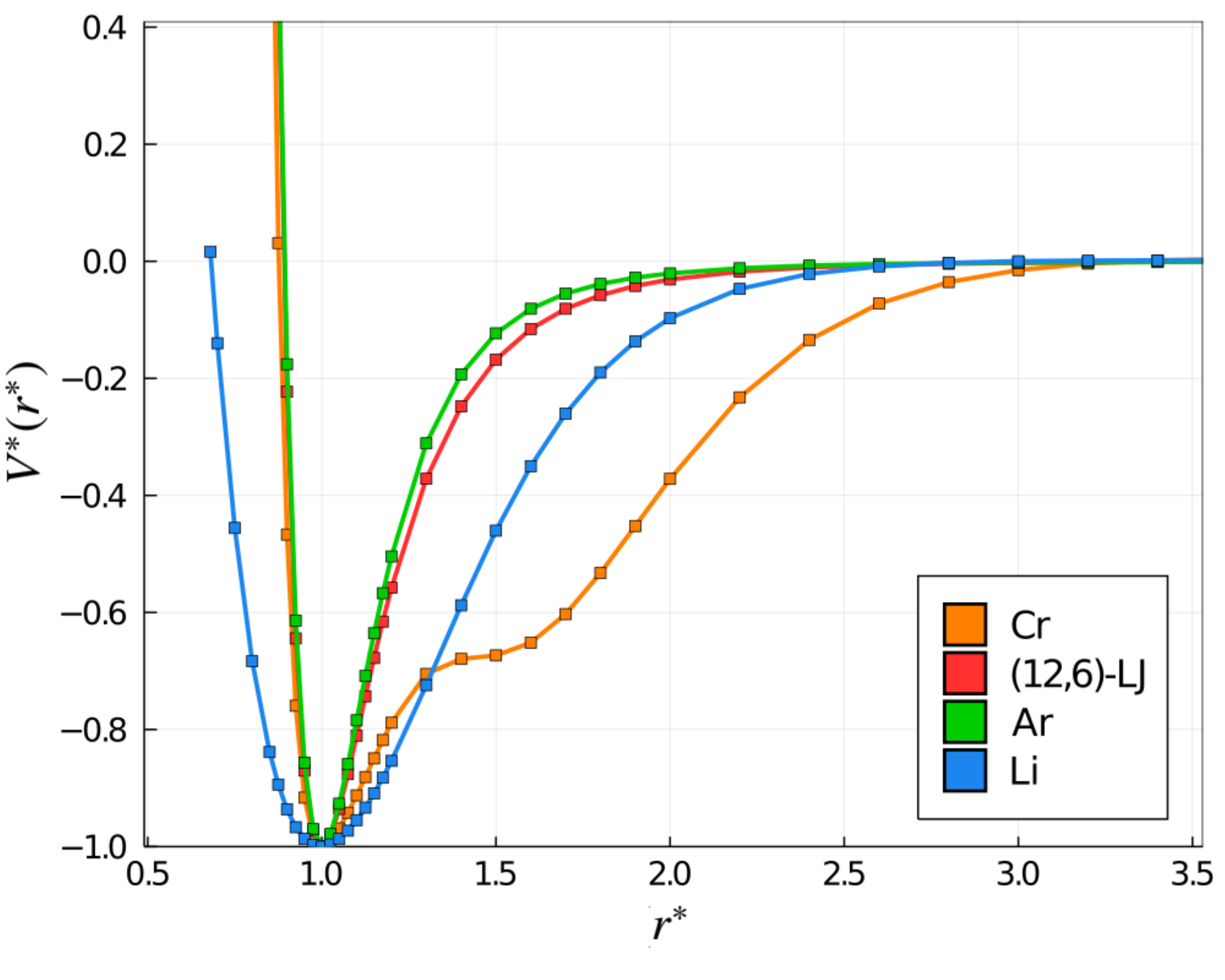}\\
  \caption{Potential energy curves $V^*(r^*)$ (in dimensionless units) for a (12-6) LJ potential, and for Ar$_2$, Li$_2$ and Cr$_2$ (see appendix).}
  \label{fig:Potcurves}
  \end{center}
  \end{figure}
  
 To underscore our argument even further the unusual potential energy curve for Cr$_2$ is considered. Here we use experimental potential values of Casey and Leopold \cite{casey1993}, but attenuated for the long range dispersion using the $C_6$ coefficient of Roos and co-workers \cite{Roos2004}, and finally fitted to an extended Lennard Jones potential potential (see appendix). This potential curve, shown in Figure \ref{fig:Potcurves}, is extremely broad and has a large dip in the medium distance range $r\in [1.3,1.7]r_e$, and therefore deviates substantially from a typical potential energy curve such as LJ or Morse \cite{vancoillie2016}. As it turns out, this potential leads to far too small distances and far too large cohesive energies for the solid state (see appendix). However, the chromium $\Delta  E^*(R^*_{\rm min},a,b,A)$ curve in Figure \ref{fig:Energies} shows that bcc remains a transition state along the distortion parameter $A$ in line with all the other two-body potentials. 
 
 We also looked at lithium, which adopts a bcc structure at normal conditions. Lithium has an extremely broad potential energy curve (see Figure \ref{fig:Potcurves}) even in the repulsive region \cite{Barakat1986}, which leads to a collapse of the crystal to a very small nearest neighbor distance (see appendix). It is clear that $N$-body forces describing correctly the confinement of the atoms in the solid state become very important here, i.e., the $N$-body expansion is not converging smoothly with increasing $N$ for metals such as lithium or chromium \cite{N-body.AH2007}. One may argue that a broad potential energy curve such as for Li$_2$ gives lower exponents for a LJ potential energy curves typical for metallic systems. It should be pointed out however, that the long range has to be correctly described and potential curves containing terms of $r^{-n}, n\le 3$ in the interaction between atoms in the solid lead to divergent series (if not analytically continued).

\section{Conclusions}
From exact lattice summations we were able to produce cohesive energies within the SHS and LJ models to computer precision. Both potentials result in an unstable bcc phase distorting toward the fcc phase or toward a phase in-between acc and bcc. The metastable bcc phase for an $(a,b)$ LJ potential occurs for unphysical potentials with very low $(a,b)$ values. The situation does not change if accurate two-body potentials are used such as for argon or chromium, the latter known to crystallize in the bcc phase. As a result, the bcc phase (at low temperatures and pressures) is stabilized only by higher than two-body forces, which have to be large enough to compete with the fcc (or hcp) structure. The mcc lattice introduced by Conway and Sloane \cite{Conway1994} is merely a point on a energetic bcc$\rightarrow$fcc downhill path. How well effective two-body potentials \cite{Johnson1988} will work for the bcc problem remains to be seen. We are currently investigating the high-pressure and temperature bcc phase stability of the bcc lattice for Lennard-Jones type of potentials. 

{\it Acknowledgements.} We acknowledge financial support by the Marsden Fund of the Royal Society of New Zealand (MAU1409).

\section*{Appendix: Lattice Sums and Their Derivatives} 

In this section we give more details about lattice sums and their derivatives, critical point analyses for the bcc structure, and potential energy curves using extended Lennard-Jones (LJ) potentials for argon, lithium and chromium.
%Method

The Gram matrix $G$ introduced in the main paper leads to the following lattice sum,  
\begin{align}
\label{latticesumf}
L(a,A)&=L(2s,A)=\mathcal{L}(s,A) = {\sum_{\vec{i}\in\mathbb{Z}^3}}^{\prime} \left( \vec{i}^\top G \vec{i} \right)^{-s} \\
\nonumber
&= {\sum_{i,j,k\in\mathbb{Z}}}^{\prime} \left(\frac{A+1}{A(i+j)^2+(j+k)^2+(i+k)^2}\right)^s,
\end{align}
with the prime indicating that the term with $\vec{i}^\top = (0,0,0)$ is not included, and $A\in [\frac{1}{3},1]$ for the cuboidal lattices considered here. These sums are important for inverse power law potentials such as LJ \cite{burrows2021a}. Here the exponent $s$ is set to $s=\frac{a}{2}$ for simplicity compared to the main paper. The lattice sums for the acc, bcc, mcc, and fcc lattices are obtained for the values $A=1/3$, $A=1/2$,  $A=1/\sqrt{2}$ and $A=1$, respectively. We split the lattice sum into two sums according to Ref.\cite{burrows2021a},
\begin{align}
\label{latticesum}
\mathcal{L}(s,A) &= \frac{(A+1)^s}{2} \left[ S_1(s,A)+S_2(s,A) \right]\\
\nonumber
&\text{with} \quad S_1(s,A)=\ {\sum_{i,j,k\in\mathbb{Z}}}^{\prime} (Ai^2+j^2+k^2)^{-s}\\
\nonumber
&\text{and} \quad S_2(s,A)= {\sum_{i,j,k\in\mathbb{Z}}}^{\prime} (-1)^{i+j+k}(Ai^2+j^2+k^2)^{-s}.
\end{align}
For the special case of $A=1$, the sum $S_1(1,s)$ represents the lattice sum for the simple cubic (sc) lattice, and the alternating sum $S_2(1,s)$ is known as the Madelung constant when $s=\frac{1}{2}$ \cite{madelung1918}. In the following, we only consider $s>\frac{3}{2}$, keeping in mind that the lattice sums are valid for all $s\in \mathbb{R}$ through analytical continuation and that $S_1(s,A)$ (and therefore $\mathcal{L}(s,A)$) has a singularity at $s=\frac{3}{2}$. 

The two lattice sums have been expanded in terms of modified Bessel functions of the second kind $K_s(x)$ in Ref.\cite{burrows2021a},
\begin{align}
\nonumber
&S_1(s,A)= a_1(s) +a_2(s) A^{1-s} \\
\label{sumpart1a}
&+a_3(s) A^{(1-s)/2} \sum_{i=1}^\infty\sum_{N=1}^\infty  c_{iN}(s) K_{s-1}\left(d_{iN}(s) \sqrt{A}\right)\\
&S_2(s,A)= b_1(s) + a_3(s) A^{(1-s)/2} \sum_{i=1}^\infty\sum_{N=0}^\infty  p_{iN}(s) K_{s-1}\left(q_{iN}(s) \sqrt{A}\right),
\label{sumpart1}
\end{align}
with the following coefficients
\begin{align}
\nonumber
&a_1(s)= 4\zeta(s)\beta(s) \quad,\quad a_2(s)= \frac{2\pi}{(s-1)}\zeta(2s-2)\\ 
\nonumber
&a_3(s)= \frac{4\pi^s}{\Gamma(s)} \quad,\quad b_1(s)= -4(1-2^{1-s})\zeta(s)\beta(s) \\
&c_{iN}(s)=r_2(N) \left(i^{-2}N\right)^{(s-1)/2}  \quad,\quad d_{iN}(s)=2\pi i \sqrt{N} \\
\nonumber
&p_{iN}(s)= (-1)^i r_2(4N+1) \left(\frac{4N+1}{2i^2}\right)^{(s-1)/2} \\
\nonumber
&q_{iN}(s)= \pi i \sqrt{8N+2} \,.
\end{align}
$\zeta(s)$ is the Riemann zeta function, $\beta(s)$ the Dirichlet beta function, and $r_2(N)$ the number of representations of number $N$ as a sum of two squares.

We are interested in the first and second derivatives, $\partial_A \mathcal{L}(s,A):=\partial \mathcal{L}(s,A)/\partial A$ and $\partial^2_A \mathcal{L}(s,A):=\partial^2 \mathcal{L}(s,A)/\partial A^2$, of the lattice sums. It was already proven directly from (\ref{latticesum}) that $\partial_A \mathcal{L}(s,A)|_{A=1/2}=0$ and $\partial^2_A \mathcal{L}(s,A)|_{A=1/2}>0$ if $s>\frac{3}{2}$ \cite{burrows2021a}. We therefore derive from eq.(\ref{latticesum}) the following expressions,
\begin{align}
\partial_A \mathcal{L}(s,A) = \frac{s}{A+1}\mathcal{L}(s,A)+\frac{(A+1)^s}{2}\left[ \partial_A S_1(s,A) + \partial_A S_2(s,A) \right]
\end{align}
and
\begin{align}
\nonumber
\partial^2_A \mathcal{L}(s,A) = -\frac{s(s+1)}{(A+1)^2}\mathcal{L}(s,A) +\frac{2s}{A+1}\partial_A \mathcal{L}(s,A) \\
+\frac{(A+1)^s}{2}\left[ \partial^2_A S_1(s,A) + \partial^2_A S_2(s,A) \right] \,.
\end{align}
The derivatives $\partial_A S_1(s,A)$, $\partial_A S_2(s,A)$, $\partial^2_A S_1(s,A)$ and $\partial^2_A S_2(s,A)$ are evaluated from the Bessel function expansions (\ref{sumpart1a}) and (\ref{sumpart1}). For this, the following relations are required,
\begin{align}
K_s(x)&=K_{s+2}(x)-\frac{2(s+1)}{x}K_{s+1}(x)\\
\nonumber
\partial_xK_s(x)&=\frac{s}{x}K_s(x) -K_{s+1} =-\frac{s}{x}K_s(x) -K_{s-1} \\
&= -\frac{1}{2}\left[ K_{s-1}(x) + K_{s+1}(x)\right] \,.
\end{align}
After some algebraic manipulations the following expressions are obtained
\begin{align}
\nonumber
\partial_A  &S_1(s,A) =- (s-1)a_2(s)A^{-s} \\
&- \frac{a_3(s)}{2}A^{-\frac{s}{2}}\sum_{i=1}^\infty\sum_{N=1}^\infty  c_{iN}(s) d_{iN}(s) K_{s}\left(d_{iN}(s) \sqrt{A}\right)
\end{align}
\begin{align}
\nonumber
\partial_A& S_2(s,A) = \\
&- \frac{a_3(s)}{2}A^{-\frac{s}{2}}\sum_{i=1}^\infty\sum_{N=0}^\infty  p_{iN}(s) q_{iN}(s) K_{s}\left(q_{iN}(s) \sqrt{A}\right)
\end{align}
\begin{align}
\nonumber
\partial^2_A& S_1(s,A) = s(s-1)a_2(s)A^{-s-1} \\
&+ \frac{a_3(s)}{4}A^{-\frac{s+1}{2}}\sum_{i=1}^\infty\sum_{N=1}^\infty  c_{iN}(s) d^2_{iN}(s) K_{s+1}\left(d_{iN}(s) \sqrt{A}\right) 
\end{align}
\begin{align}
\nonumber
\partial^2_A& S_2(s,A) =  \\
&\frac{a_3(s)}{4}A^{-\frac{s+1}{2}}\sum_{i=1}^\infty\sum_{N=0}^\infty  p_{iN}(s) q^2_{iN}(s) K_{s+1}\left(q_{iN}(s) \sqrt{A}\right) \,.
\end{align}
The Bessel function sums are fast converging, therefore making the evaluation of lattice sums and their derivatives to computer precision attainable within less than a second on a modern laptop computer \cite{ProgramJones}.
  
\subsection*{Critical Points for the bcc structure} 
For the following, we set $a$ to $2s$, making the lattice sum $L(a,A)=L(2s,A)$, which is more convenient for the LJ potential.
Figure 1 shows the lattice sums and their second derivative for $A=\frac{1}{2}$ (bcc lattice) as a function of the exponent $a$. It is clear that  $\partial^2_A L(A,a)|_{A=\frac{1}{2}}$ has a peculiar form with a minimum at $a=5.52534$ and a maximum at $a=12.57676$, this becomes important in the discussion of the bcc stability for Lennard-Jones systems detailed below. However, it is illustrative to evaluate the minimum distance derivatives $\partial^n_A R^*_{\rm min}(a,b,A)$ for $A=\frac{1}{2}$ and $n=1,2$ (using dimensionless quantities as discussed in the main paper).
 \begin{figure}[htbp]
  \begin{center}
  \includegraphics[scale=0.225]{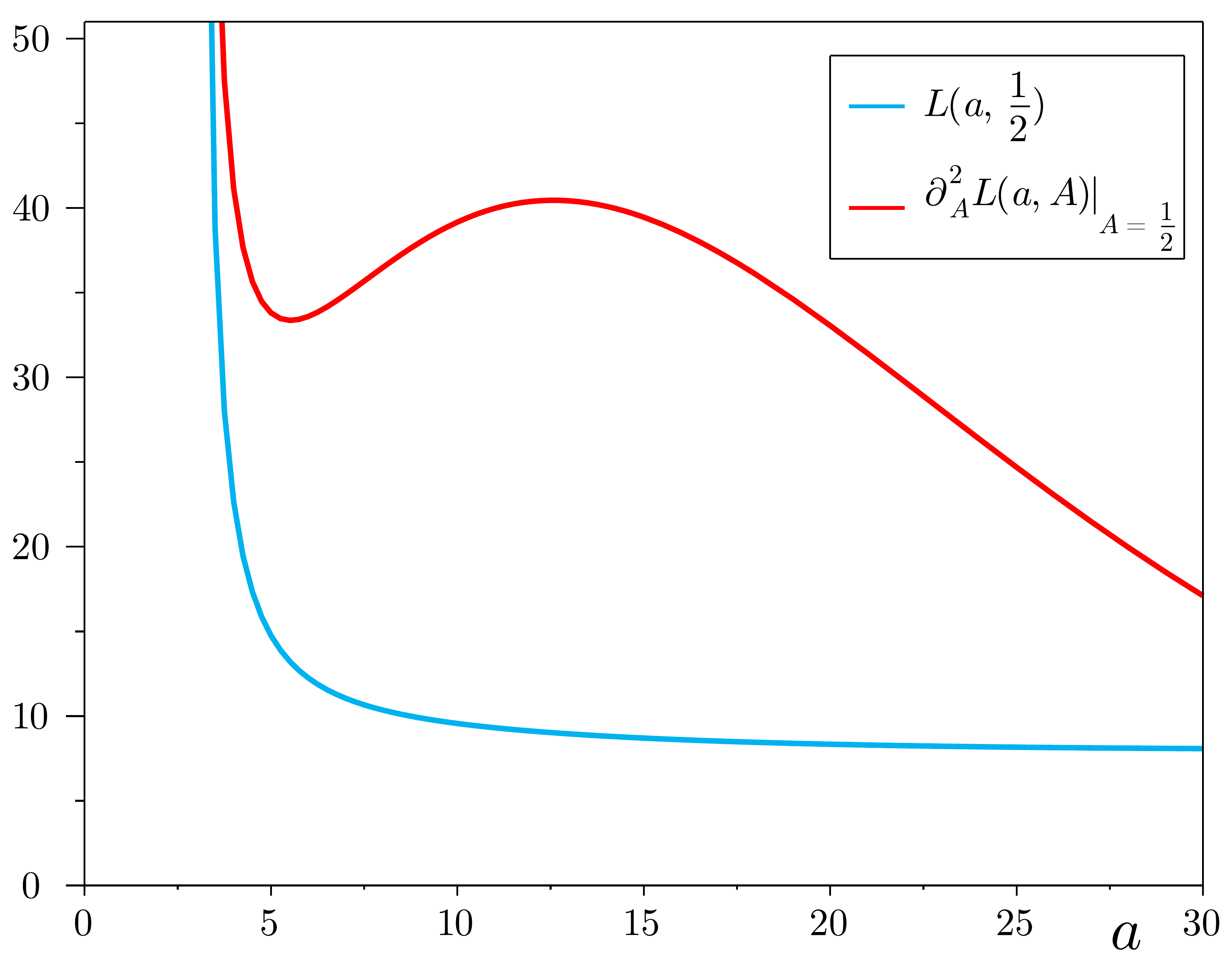}\\
  \caption{Lattice sums $L(a,\frac{1}{2})$ and $\partial^2_A L(a,A)|_{A=\frac{1}{2}}$ (bcc lattice) as a function of the exponent $a$. Note that $\partial_A L(a,A)|_{A=\frac{1}{2}}=0$ for all $a$ values.}
  \label{fig:LatticeSums}
  \end{center}
  \end{figure}
The minimum distance for the cuboidal lattices is given by 
\begin{equation} \label{eq:mindist}
 R^*_{\rm min}(a,b,A)=\left( \frac{L(a,A)}{L(b,A)}\right)^{\tfrac{1}{a-b}}\,,
\end{equation}
and is shown in Figure \ref{fig:Distances} for various $(a,b)$ combinations.
\begin{figure}[hb]
  \begin{center}
  \includegraphics[scale=0.24]{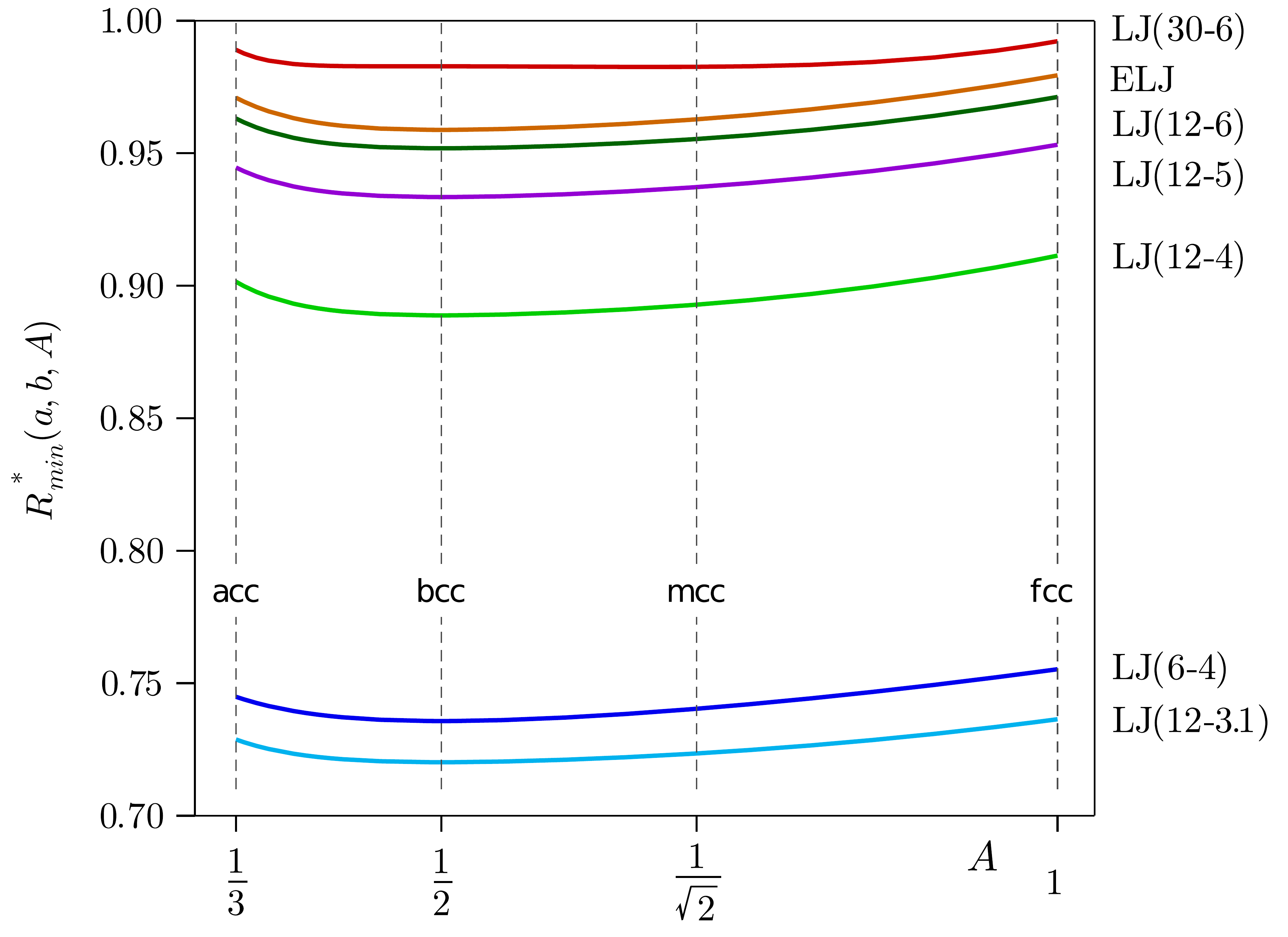}
  \caption{Minimum distance $R^*_{\rm min}(a,b,A)$ for various $(a,b)$ LJ potentials and for the ELJ potential (taken from Ref.\cite{Schwerdtfeger-2016}) dependent on the lattice parameter $A$.}
  \label{fig:Distances}
  \end{center}
  \end{figure}
The first derivative for $a>b>3$ is given by
\begin{equation} \label{eq:mindistderiv}
\partial_A R^*_{\rm min}(a,b,A)=\frac{R^*_{\rm min}(a,b,A)}{a-b}\left( \frac{\partial_A L(a,A)}{L(a,A)} - \frac{\partial_A L(b,A)}{L(b,A)}\right) \,,
\end{equation}
which for the bcc lattice ($A=\frac{1}{2}$) is zero because $\partial^A L(a,A)|_{A=\frac{1}{2}}=0$ identically for all values of $a>3$  \cite{burrows2021a}. The second derivative evaluated at $A=\frac{1}{2}$ is given by
\begin{align} \label{eq:mindistderiv2}
\partial^2_A R&^*_{\rm min}(a,b,A)|_{A=\frac{1}{2}}=\\
\nonumber
&\left[\frac{R^*_{\rm min}(a,b,A)}{a-b}\left( \frac{\partial^2_A L(a,A)}{L(a,A)} - \frac{\partial^2_A L(b,A)}{L(b,A)} \right)\right]_{A=1/2} \,.
\end{align}
Evaluating the expression in parentheses in (\ref{eq:mindistderiv2}) shows that $R^*_{\rm min}(a,b,A)$ has a minimum at $A=\frac{1}{2}$ if $a<14.17598$. For values $a>14.17598$ we have a certain range of $b$ values where $R^*_{\rm min}(a,b,A)$ becomes a shallow maximum as is the case for the (30,6) LJ potential shown in Figure \ref{fig:Distances}.

\begin{figure}[htbp]
  \begin{center}
  \includegraphics[scale=0.225]{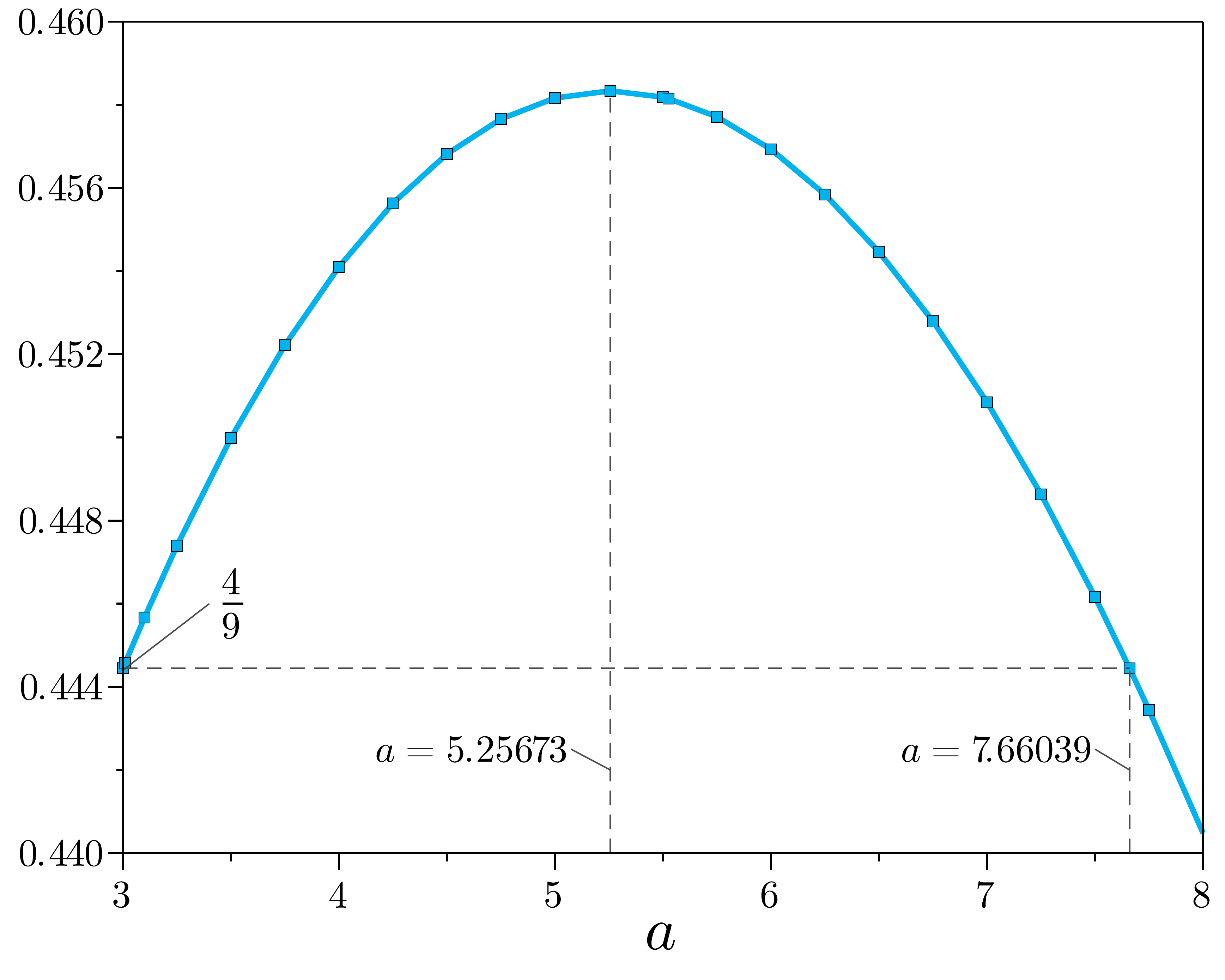}\\
  \caption{Lattice sum ratio $\frac{\partial_A^2 L(a,A)|_{A=\frac{1}{2}}}{aL(a,\frac{1}{2})}$ against the exponent $a$ at lattice parameter $A=\frac{1}{2}$.}
  \label{fig:2ndderiv2}
  \end{center}
  \end{figure}

In a similar way we evaluate the cohesive energy for an $(a,b)$ LJ potential at $R^*_{\rm min}(a,b,A)$,
\begin{align} \label{eq:EcohLJmn}
\nonumber
E^*&(R^*_{\rm min},a,b,A) =\\
\nonumber
&\frac{1}{2(a-b)}\left[bL(a,A)\left(\frac{L(b,A)}{L(a,A)}\right)^{\tfrac{a}{a-b}} - aL(b,A)\left(\frac{L(b,A)}{L(a,A)}\right)^{\tfrac{b}{a-b}}\right]\\
&=-\frac{1}{2} \left[ \frac{L(b,A)^{1/b}}{L(a,A)^{1/a}}\right]^\frac{ab}{a-b} \,.
\end{align}
%&=-\frac{1}{2} \left[ \frac{L(b,A)^\tfrac{1}{b}} {L(a,A)^\frac{1}{b} } \right]^\frac{ab}{a-b}
The first and second derivatives are evaluated as,
\begin{align} \label{eq:EcohLJmnd1}
&\partial_A E^*(R^*_{\rm min},a,b,A) =\\
\nonumber
&E^*(R^*_{\rm min},a,b,A)\frac{ab}{(a-b)}\left[ \frac{1}{b}\frac{\partial_A L(b,A)}{L(b,A)} - \frac{1}{a}\frac{\partial_A L(a,A)}{L(a,A)} \right]\\
\nonumber
\textrm{and}\\
&\partial_A^2 E^*(R^*_{\rm min},a,b,A) = \frac{\left\{ \partial_A E^*(R^*_{\rm min},a,b,A)\right\} ^2}{E^*(R^*_{\rm min},a,b,A)}\\
\nonumber
&+\frac{ab}{a-b} E^*(R^*_{\rm min},a,b,A)\bigg[  \frac{1}{b} \frac{\partial_A^2 L(b,A)}{L(b,A)} -  \frac{1}{a} \frac{\partial_A^2 L(a,A)}{L(a,A)} \\
\nonumber
&- \frac{1}{b}\frac{\left\{\partial_A L(b,A)\right\}^2}{L(b,A)^2} +  \frac{1}{a}\frac{\left\{\partial_A L(a,A)\right\}^2}{L(a,A)^2}\bigg] \,.
\end{align}
The first derivative is zero for the bcc lattice ($A=\frac{1}{2}$) because $\partial^A L(a,A)|_{A=\frac{1}{2}}=0$ as mentioned above. This makes the bcc point strictly an extremum along the $A$ coordinate for any $(a,b)$ combination of the LJ potential. The second derivative evaluated at $A=\frac{1}{2}$ gives
\begin{align} \label{eq:EcohLJmnd2}
\partial_A^2 &E^*(R^*_{\rm min},a,b,A)|_{A=\frac{1}{2}} = \\
\nonumber
&\frac{ab}{a-b} E^*(R^*_{\rm min},a,b,\tfrac{1}{2})\left[  \frac{1}{b} \frac{\partial_A^2 L(b,A)|_{A=\frac{1}{2}}}{L(b,\frac{1}{2})} -  \frac{1}{a} \frac{\partial_A^2 L(a,A)|_{A=\frac{1}{2}}}{L(a,\frac{1}{2})} \right] \,.
\end{align}
Hence, the bcc instability can be a maximum or a (metastable) minimum depending on the sign of the expression in the square brackets. The transition to a metastable phase occurs at
\begin{align} \label{eq:phasetrans}
 \frac{\partial_A^2 L(b,A)|_{A=\frac{1}{2}}}{bL(b,\frac{1}{2})} =  \frac{\partial_A^2 L(a,A)|_{A=\frac{1}{2}}}{aL(a,\frac{1}{2})} \,,
\end{align}
with $b<a$. For the singularity at $a=3$ we get from computation,
\begin{align} \label{eq:limit}
\lim_{a\rightarrow\ 3}\frac{\partial_A^2 L(a,A)|_{A=\frac{1}{2}}}{aL(a,\frac{1}{2})} = \frac{4}{9} \,,
\end{align}
which is shown on Figure \ref{fig:2ndderiv2}. This can be proven using a Laurent expansion around the simple pole at $a=3$ \cite{burrows2021a}, 
\begin{align} \label{eq:Laurent}
L(A;s) = \frac{2c_{-1}(A)}{a-3}+c_0(A)+\sum_{n=1}^\infty 2^{-n}c_n(A) (a-3)^n
\end{align}
with
\begin{align} \label{eq:Laurentcoef}
c_{-1}(A) = \pi\sqrt{\frac{(A+1)^3}{A}}\quad {\rm and} \quad \partial^2_A c_{-1}(A) = \frac{3\pi}{4A^2\sqrt{A(A+1)}} \,.
\end{align}
This gives
\begin{align} \label{eq:Laurentquot}
\frac{\partial_A^2 L(a,A)}{L(a,A)}=\frac{ \partial^2_A c_{-1}(A)}{c_{-1}(A)}+\mathcal{O}(a-3)=\frac{3}{4A^2(A+1)^2}+\mathcal{O}(a-3)
\end{align}
which results in (\ref{eq:limit}) for  $a=3$ and $A=\frac{1}{2}$.

From this limit it is clear that a metastable minimum can only exist if $a<a_\text{MS}=7.66039$, but with a limited range of small $b$ values evident from (\ref{eq:phasetrans}) and Figure \ref{fig:2ndderiv2}. The maximum of the curve shown in Figure \ref{fig:2ndderiv2} is at $a_\text{max}$=5.25673, for which all $b<a<a_\text{max}$ values result in a metastable state. We note that the curve in Figure \ref{fig:2ndderiv2} is almost (but not quite) symmetric around the maximum. This makes the phase transition line from the unstable to the metastable bcc phase almost linear in the $(a,b)$ plane.

Next, we consider the Einstein frequency of a single atom of mass $M$ moving in the field of other atoms (in atomic units) for an $(a,b)$ LJ potential \cite{Smits2020a},
\begin{align} \label{eq:EZPELJn-m}
\omega_E&(R,a,b,A) =
\nonumber
 \frac{1}{3r_e}\sqrt{\frac{3\epsilon}{M}}\sqrt{\frac{ab}{a-b}}\left(\frac{r_e}{R}\right)^{\frac{a}{2}+1}\\
&\times \left[ (a-1) L(a+2,A) -(b-1) L(b+2,A) \left(\frac{R}{r_e}\right)^{a-b}\right]^{\frac{1}{2}}.
\end{align}
It is clear that $\omega_E(R)$ describes the instability of lattice by moving a single atom as opposed to a collective movement of several atoms in the lattice. However, at $R_\text{min}$ we always arrive at $\omega_E(R_\text{min})>0$ for a finite mass $M$. To prove this we have to show that the term in the square brackets stays positive at $R_\text{min}$ for a fixed $A$ value, that is
\begin{equation}
\frac{L(a+2,A)}{L(b+2,A)}>\frac{(b-1)L(a,A)}{(a-1)L(b,A)} \,.
\end{equation}
As $a>b$ it suffices to show that 
\begin{equation}
\nonumber
\frac{L(a+2,A)}{L(b+2,A)}\ge\frac{L(a,A)}{L(b,A)} \,,
\end{equation}
or more generally
\begin{equation}
\label{inequality}
\frac{L(b,A)}{L(b+h,A)}\ge\frac{L(a,A)}{L(a+h,A)} 
\end{equation}
for any $h>0$ and $a>b>3$. The proof is given further below in the last section.

\subsection*{Extended Lennard-Jones Potentials for Li$_2$, Ar$_2$ and Cr$_2$} 

The extended Lennard-Jones potential is defined by
 \begin{equation} \label{eq:extLJ}
V_{\rm ELJ}(r,c_n) =\sum_{n=1}^{n_\textrm{max}} c_nr^{-a_n} \quad {\rm with} \quad \sum_{n=1}^{n_\textrm{max}} c_nr_e^{-a_n}=-D_e \,.
\end{equation}
It then follows that the cohesive energy for an extended Lennard-Jones potential becomes,
\begin{equation} \label{eq:extLJ}
E_{\rm ELJ}(R,c_n,A) =\frac{1}{2}\sum_{n=1}^{n_\textrm{max}} c_nL(a_n,A)R^{-a_n}
\end{equation}
with $R$ being the nearest neighbor distance in the solid. The corresponding parameters for the potential energy curves in reduced units, $V^*(r^*)$, for Ar$_2$, Li$_2$ and Cr$_2$ are listed in Table 1. For Ar$_2$ the extended LJ potential from Ref.\cite{Schwerdtfeger-2016} has been converted to dimensionless units for this work ($r^*=r/r_e$, $V^*(r)=V^*(r)/D_e$, from which follows that $r^*_\text{min}=1$ and $V^*_\text{min}=-1$). 

\begin{table}[htb]
\centering
\caption{Potential parameters for the Ar, Li and Cr dimers obtained from a least-squares fit to the (a) analytical form of Cybulski and Toczyłowski for Ar$_2$ \cite{CybulskiToczylowski1999,Schwerdtfeger-2016}, (b) exp. determined potential of Barakat et al. \cite{Barakat1986}, and (c) exp. determined potential of Casey and Leopold \cite{casey1993} as described in detail in the text. Dimensionless units are used. For Li$_2$ and Cr$_2$ the potential parameters are only valid for the region $V^*(r^*)<0$.}
\label{tab:2bodyfitting}
\begin{tabular}{r|l|r|r|l|r}
 \hline  \hline
 $n$ & $a_n$  & $c_n$ & $n$ & $a_n$  & $c_n$   \\
\hline
Ar &&&&&\\
1  	& 6   & -2.112319339 &  2 & 8 & 7.126409258  \\
3  	& 10   & -21.30053312 &  3 & 12 & 24.42390886  \\
5  	& 14 & -10.89025935 &  6 & 16 & 1.752793693  \\
\hline
Li &&&&&\\
1  	& 6   &-2.185099402  &  2  & 8   &1588.743093   \\
3 	& 9   &-13096.66094  &  4  & 10 &44937.24250  \\
5 	& 11 &-85547.67477  &  6  & 12 &100055.5130   \\
7 	& 13 &-74450.14624  &  8  & 14 &35150.12854   \\
9 	& 15 &-10264.39581  & 10 & 16 &1744.182010   \\
11 	& 17 &25.87885791  & 12 & 18 &-237.6273332   \\
13 	& 19 &114.6392978  & 14 & 20 &-18.63705649   \\
\hline
Cr &&&&&\\
1  	& 6   & -15.20122639 &  2  & 8   & 13471.86476  \\
3 	& 9   & -124591.4050 &  4  & 10 & 464698.3696  \\
5 	& 11 & -888076.6787 &  6  & 12 & 854878.9650  \\
7 	& 13 & -190568.3900 &  8  & 14 &  -441981.1016 \\
9 	& 15 & 487340.5171 & 10 & 16 &  -209652.8384 \\
11 	& 17 & 34494.89857 & 12 & 18 &  0.000016589 \\
\hline  \hline
\end{tabular}
\end{table}

For Cr$_2$ we took the potential curve from experimental data of Casey and Leopold, who obtained the potential energy curve $V(r)$ from vibrational data through the RKR method \cite{casey1993}. This potential curve only describes the medium range of the potential energy curve. We therefore attenuated the long range by matching the last point $R_\text{max}=$3.35 \AA~ to a $-C_6/r^{-6}$ dispersion curve. Finally, the points are used to fit an inverse power potential (extended Lennard-Jones) to the potential energy curve fixing the Van der Waals coefficient to $C_6=$800 a.u. according to Roos and co-workers \cite{Roos2004}. Because of the peculiar shape of the Cr$_2$ potential energy curve the fit was rather difficult to achieve, but is accurate enough ($R^2$=0.9984) for the discussion of the bcc instability. The potential energy curve for Cr$_2$ was then converted to dimensionless units. For the ELJ form we obtain $E^*=$24.0 and 23.3 for the fcc and bcc structures respectively. These values are unusually large, but perhaps not surprising given the broad potential energy curve of Cr$_2$. In fact, using the original potential energy curve we obtain a nearest neighbor distance for bcc chromium of $R_{\rm min}= 1.479$ \AA, just above the hard sphere radius of the diatomic potential energy curve with $\sigma=1.467$ \AA, and a cohesive energy $E_{coh}= 33.6$ eV. This is in stark disagreement with the experimental values of $R_{\rm min}=2.52$ \AA~ and $E_{\rm coh}=4.1$ eV \cite{kittel1996}. It clearly demonstrates that the direct use of potential curves from the free unconfined diatomic is not useful to describe the solid state of metals as the many-body expansion is not converging fast and smoothly.

We briefly discuss lithium. For Li$_2$ we used the Rydberg-Klein-Rees (RKR) potential curve of Barakat et al. \cite{Barakat1986} and fixed the Van der Waals coefficient $C_6=$1408 a.u. \cite{Gould2016}. For the fit to an extended LJ potential we obtained with an $R^2$ value of 0.99997, but only by including terms up to $1/r^{20}$. However, the situation here is even worse compared to chromium as the Li$_2$ potential energy curve is so broad in both the repulsive and attractive region that crystal optimizations entered the repulsive wall well below the hard-sphere radius of $\sigma=1.822$ \AA, where our extended LJ potential is not accurate anymore. In general, a fit to an extended LJ form works reasonably well for the whole distance region if it deviates not too much from an ideal ($a,b$)-LJ potential, which is certainly not the case for Li$_2$. In fact, if we optimize the exponents $a,b$ for the LJ potential we get $a\approx b<3$ left of the singularity at $b=3$ and therefore an unphysical result. Using the far more accurate extended Morse potential by LeRoy and co-workers \cite{LeRoy2011}, which correctly describes the repulsive region, we obtain from crystal optimizations \cite{ProgramSamba} a nearest neighbor distance of $R_{\rm min}= 0.21$ \AA~ and a cohesive energy of $E_{\rm coh}= 9.2\times10^3$ eV for bcc lithium. This can be best described as a collapse of the crystal to small internuclear distances with large overbinding, and clearly demonstrates that many-body forces in a confined bulk system cannot be neglected.

\subsection*{Proof of inequality eq.(\ref{inequality})} 

%This is indeed fulfilled as for $a>b>3$, $L(b,A)$ is a convex and monotonically decreasing function in $b$ with $L(b,A)\ge1$. Therefore the bcc instability comes from a collective motion of the atoms.
A function $g(x)$ is said to be logarithmically convex on an interval if \mbox{$g(x)>0$} and  $\log g(x)$ is convex on the interval.
It can be shown that the sum of logarithmically convex functions is logarithmically convex, e.g., see~\cite[p. 19]{Roberts1973}.
It follows that the lattice sum $L(x,A)$ is a logarithmically convex function of~$x$ because it is a sum of terms of the form $n^{-x}$,
each of which is logarithmically convex.

Now suppose that $f(x)$ is a convex function, and $x_1$, $x_2>0$. By applying the definition of convexity to the interval $[0,x_1+x_2]$ we have
$$
f(x_1) \leq \frac{x_2}{x_1+x_2}f(0) + \frac{x_1}{x_1+x_2}f(x_1+x_2),
$$
while interchanging $x_1$ and $x_2$ gives
$$
f(x_2) \leq \frac{x_1}{x_1+x_2}f(0) + \frac{x_2}{x_1+x_2}f(x_1+x_2).
$$
Adding the inequalities gives
\begin{equation}
\label{P1}
f(x_1)+f(x_2) \leq f(0)+f(x_1+x_2).
\end{equation}
Incidentally, it can be shown from this using mathematical induction that
$$
f(x_1)+f(x_2)+\cdots +f(x_n) \leq (n-1)f(0)+ f(x_1+x_2+\cdots+x_n),
$$
a result known as Petrovi\'{c}'s inequality, e.g., see~\cite[p. 22]{Mitrinovic1970}, \cite{Petrovic1932}. We shall only require the case~$n=2$ as given by~\eqref{P1}.

Suppose $a>b$, $h>0$, and $g(x)$ is a convex function for $x\geq b$.
Let $f(x) = g(x+b)$ and take $x_1=h$ and $x_2=a-b$. Then Petrovi\'{c}'s inequality~\eqref{P1} gives
$$
f(h)+f(a-b) \leq f(0)+f(a-b+h).
$$
This implies
$$
g(b+h)+g(a) \leq g(a+h)+g(b)
$$
which is equivalent to
$$
g(b+h)-g(b) \leq g(a+h)-g(a).
$$
It follows that if $G(x)$ is logarithmically convex, then
$$
\log G(b+h)-\log G(b) \leq \log G(a+h)-\log G(a).
$$
This can be rearranged to give
$$
\frac{G(b+h)}{G(b)} \leq \frac{G(a+h)}{G(a)},
$$
which is exactly the inequality we seek for the lattice sums.

\bibliography{references}
\end{document}